# DISSIPATIVE TUNNELING IN STRUCTURES WITH QUANTUM DOTS AND QUANTUM MOLECULES

**Yu.I. Dahnovsky**[1], **V.D. Krevchik**[2, 3], **M.B. Semenov**[2, 3], **K. Yamamoto**[4], **V.Ch. Zhukovsky**[5], **A.K. Aringazin**[3, 6], **E.I. Kudryashov**[2], **V.G. Mayorov**[2]

[1)] Department of Physics & Astronomy/3905, 1000 E. University Ave., University of Wyoming, Laramie, WY 82071, USA
[2)] Physics Department, Penza State University, Penza, Russia, physics@diamond.stup.ac.ru
[3)] Institute for Basic Research, P.O. Box 1577, Palm Harbor, FL 34682, USA
ibr@verizon.net
[4)] Research Institute of International Medical Center, Japan
[5)] Department of Theoretical Physics, Moscow State University, Moscow, Russia
[6)] Institute for Basic Research, Eurasian National University, Astana 010008, Kazakhstan
aringazin@mail.kz

The problem of tunneling control in systems “quantum dot – quantum well” (as well as “quantum dot – quantum dot” or quantum molecule) and “quantum dot – bulk contact” is studied as a quantum tunneling with dissipation process in the semiclassical (instanton) approximation. For these systems temperature and correlation between a quantum dot radius and a quantum well width (or another quantum dot radius) are considered to be control parameters. The condition for a single electron blockade is found in the limit of quantum dot. The criteria for an extreme tunneling current in quantum molecules are also presented. The tunnel probability for systems under investigation is analytically obtained as well.

# 1. Introduction

The problem of an electron tunnel transfer in nanostructures has been of great interest in the recent years [1-6]. For example, the authors of Ref. [13] observed individual tunnel events of a single electron between a quantum dot and a reservoir, using a nearby quantum point contact as a charge meter. Also, the report on realization of few-electron double quantum dots defined in a two-dimensional electron gas by means of surface gates on top of a GaAs/AlGaAs heterostructure can be found in Ref. [14]. The experiments demonstrate that this quantum dot circuit can serve as a good starting point for a scalable spin, the qubit system. Microwave spectroscopy of a quantum-dot molecule has been also represented in Ref. [15].

The fundamental and applied significance of this problem is based, first of all, on an extremely high sensitivity of the tunneling probability to an electron energy spectrum, a confinement potential profile, and external field parameters. The latter provides an additional degree of freedom for a possible control of different conjugated nanostructures. In this work we primarily interested in "quantum dot - quantum well" (or a quantum molecule) and "quantum dot - bulk contact" structures.

In particular, we calculate the tunneling probability of an electron with an exponential accuracy in the framework of the theory developed in Refs. [5, 7, 8], where tunneling structures with quantum dots are considered as chemical reaction systems. Such a similarity opens up a possibility to employ a powerful apparatus of quantum tunneling with dissipation [5, 9, 10, 11] earlier developed in the theory of tunneling chemical reactions where the interaction of electron with an environment is essential.

It is well known from Quantum Mechanics that quantum tunneling occurs when the characteristic width of a quantum well is much greater than a de Broglie wavelength of a tunneling particle. In the instanton or semiclassical approximation this criterion yields [5, 7, 8]:

$$
\begin{cases}
R >> \dfrac{\hbar}{\left(2-\sqrt{3}\right)\sqrt{2m^{*}U_0}} & (1) \\
R >> \dfrac{\hbar}{\sqrt{8m^{*}k_B T}} & (2)
\end{cases}
$$

where $R$ is a quantum dot radius, $U_0$ is a barrier height, and $m^*$ is an effective electron mass. Inequality (2) provides the criterion for the instanton – antiinstanton pairs rare gas approximation [5, 7, 8], where a subsequent tunneling event is considered to be independent on a previous one. For InSb, the validity conditions (1) and (2) can be satisfied simultaneously if $T \approx 50\,K$ .

In this work we study the following two different systems:

1) "*quantum dot – quantum well*" (or *quantum molecule*) to be modeled by a double-oscillator ("nondecayble") system interacting with a bath, and

2) "*quantum dot – bulk contact*" to be modeled by an oscillator well cut off by a vertical wall (decayable potential model).

First model can describe tunnel transfer in quantum molecules (i.e., interacting quantum dots) or separate quantum dots, connected by the tunnel with quantum well. Second model (the decayble one) can represent tunneling transfer in a system with separate quantum dots on surface of bulk matrix (or contact).

To study the electron tunneling we employ the dissipative quantum tunneling in adiabatic chemical reactions in the semiclassical approximation [12]. It is known that adiabatic tunneling reactions are defined a Landau-Zener parameter is large, i.e.

$$
\frac{\Delta^2}{\hbar u\left|F_2 - F_1\right|} >> 1,
$$

where $\Delta$ is the electron transition matrix, $u$ is the particle velocity, and $F_{1,2}$ are the forces at the terms intersection point. For the reactive system, potential energy surfaces in the initial and final states are modeled by a set of independent shifted oscillators with the same frequency [5, 7, 8]:

$$U_i = \sum_{i=1}^{N} \frac{1}{2} \omega_{0i}^{2} \left( x_i + x_{0i} \right)^2 , \; U_f = \sum_{i=1}^{N} \frac{1}{2} \omega_{0i}^{2} \left( x_i - x_{0i} \right)^2 - \Delta \mathrm{I} \quad (3)$$

As shown in Ref. [9], Hamiltonian with such a potential energy can be presented as one-dimensional tunneling along some tunnel coordinate $y_1$ linearly coupled with harmonic bath degrees of freedom,

$$\hat{H} = \frac{p_1^{2}}{2} + v_1(y_1) + y_1 \sum_{\alpha=2}^{N} C_\alpha y_\alpha + \frac{1}{2} \sum_{\alpha=2}^{N} \left( p_\alpha^{\ 2} + \omega_\alpha^{\ 2} y_\alpha^{\ 2} \right), \quad (4)$$

where $C_\alpha$ is a coupling constant for α-th mode. For such a system, the tunneling probability per unit time is determined as follows [9, 10]:

$$\Gamma = 2T \frac{\operatorname{Im} Z}{\operatorname{Re} Z} \quad (5)$$

where

$$Z = \prod_\alpha \int Dy_1 \int Dy_\alpha \exp\left[ -S\{y_1; y_\alpha\} \right]$$

(6)

is a partition function of the system that can be represented as a path integral with the following boundary conditions:

$$y_\alpha(-\beta/2) = y_\alpha(\beta/2),$$

where $\beta \equiv T^{-1}$ or $\beta \equiv \frac{\hbar}{kT}$ (we have used the units where $\hbar = 1$ and $k = 1$, and all the oscillator masses are equal to 1). Here $S$ denotes a multidimensional Euclidean action for the whole system. The imaginary part $\operatorname{Im} Z$ determines the decay of energy states in the initial well. Obviously, dissipation should be strong enough to avoid coherent oscillation for the electron between two wells. In the instanton approximation, the Euclidean action yields [7, 9, 10]

$$S\{y_1\} = \int_{-\beta/2}^{\beta/2} d\tau \left[ \frac{1}{2} \dot{y}_1^{\ 2} + v(y_1) + \frac{1}{2} \int_{-\beta/2}^{\beta/2} d\tau' K\left( \tau - \tau' \right) y_1(\tau) y_1(\tau') \right], \quad (7)$$

$$v(y_1) = v_1(y_1) - \frac{1}{2} \sum_{\alpha=2}^{N} \frac{C_\alpha^{\ 2}}{\omega_\alpha^{\ 2}} y_1^{\ 2}, \quad (8)$$

$$\zeta_n = \nu_n^{\ 2} \sum_{\alpha=2}^{N} \frac{C_\alpha^{\ 2}}{\omega_\alpha^{\ 2}\left(\omega_\alpha^{\ 2} + \nu_n^{\ 2}\right)}, \tag{9}$$

where $\nu_n \equiv 2\pi nT$ is the Matsubara's frequency.

In Secs. 2 and 3, we consider two different systems, a "quantum dot - quantum well" system (or quantum molecule) and a "quantum dot – bulk contact" system. In Sec. 4, we summarize the results.

## 2. "Quantum dot –quantum well" or quantum molecule system

As mentioned in Introduction, a tunneling potential along the tunneling coordinate can be presented in the following way [9] (see Fig. 1):

$$v(q) = \frac{1}{2}\omega_0^{\ 2}\left(q+q_0\right)^2 \theta(-q) + \left[\frac{1}{2}\omega_0^{\ 2}\left(q-q_1\right)^2 - \Delta I\right]\theta(q), \tag{10}$$

$$q = y_1 + \frac{\Delta I}{2\lambda},\ \omega_0^{\ 2} = \omega_1^{\ 2} - \sum_{\alpha=2}^{N}\frac{C_\alpha^{\ 2}}{\omega_\alpha^{\ 2}},\ q_0 = \frac{\lambda}{\omega_0^{\ 2}} - \frac{\Delta I}{2\lambda},\ q_1 = \frac{\lambda}{\omega_0^{\ 2}} + \frac{\Delta I}{2\lambda},$$

$$\lambda^2 = \sum_{i=1}^{N} \omega_{0i}^{\ 4} x_{0i}^{\ 2}$$

To calculate the instanton action, it is necessary to find an instanton, i.e. a trajectory that minimizes the Euclidean action $S(q)$. This trajectory can be found from the following equation of motion that includes the dissipation to the bath:

$$-\ddot{q}_B(\tau) + \frac{\partial v(q_B)}{\partial q_B} + \int_{-\beta/2}^{\beta/2} d\tau' K(\tau-\tau') q_B(\tau') = 0, \quad q_B(\tau) = q_B(\tau+\beta) \tag{11}$$

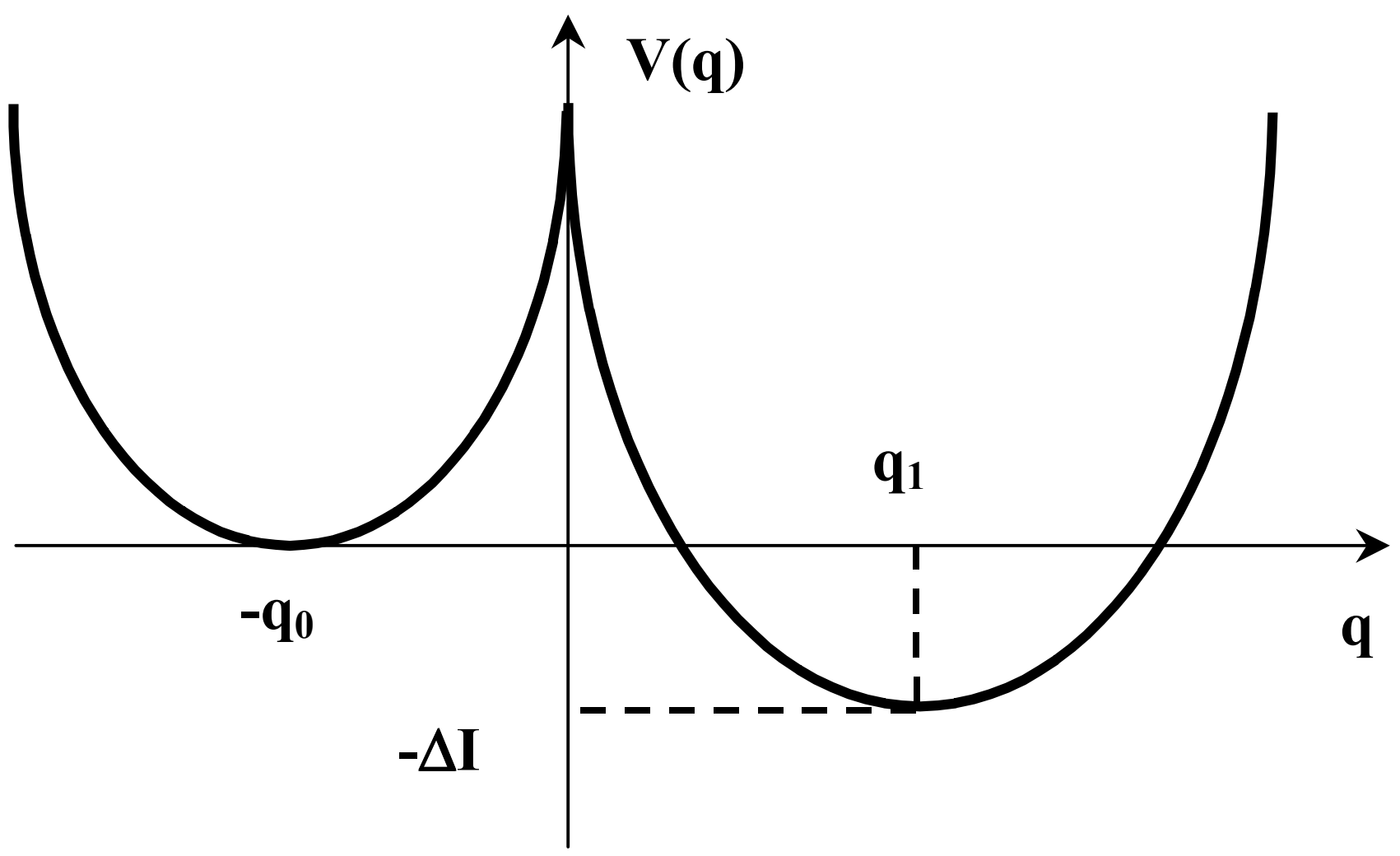

**Fig. 1.** Potential energy along the reaction coordinate.

A solution of the equation (11) is sought as a periodic function

$$q_B(\tau) = \beta^{-1} \sum_{n=-\infty}^{\infty} q_n \exp\left(i\nu_n \tau\right), \tag{12}$$

Inserting Eq. (12) into Eq.(11), one finds that

$$q_B(\tau) = -q_0 + \frac{2\left(q_0 + q_1\right)\tau_0}{\beta} + \frac{2\omega_0^{\ 2}\left(q_1 + q_0\right)}{\beta} \sum_{n=1}^{\infty} \frac{\sin\nu_n\tau_0 \cdot \cos\nu_n\tau}{\nu_n\left(\nu_n^{\ 2} + \omega_0^2 + \zeta_n\right)}. \tag{13}$$

The instanton action can be easily found by inserting the instanton (13) into the expression for the action (7),

$$S_B = 2\omega_0^{\ 2}\left(q_0 + q_1\right) q_0 \tau_0 - \frac{2\omega_0^{\ 2}\left(q_0 + q_1\right)^2 \tau_0^{\ 2}}{\beta} -$$

$$-\frac{4\omega_0^{\ 4}\left(q_0 + q_1\right)^2}{\beta} \sum_{n=1}^{\infty} \frac{\sin^2 \nu_n \tau_0}{\nu_n^{\ 2}\left(\nu_n^{\ 2} + \omega_0^{\ 2} + \zeta_n\right)}. \tag{14}$$

In case of interaction with a single mode $\omega_L$, the instanton action (14) can be found analytically [9],

$$2S = \left(q_1 + q_0\right)\left(3q_0 - q_1\right)\omega^2 \tau_0 - \frac{4\omega^2\left(q_0 + q_1\right)^2 \left(\tau_0\right)^2}{\beta} -$$

$$
-\frac{\omega^2 (q_0+q_1)^2}{2\tilde{\gamma}}\left\{\frac{(\omega^2-\tilde{x}_2)}{\sqrt{\tilde{x}_1}}\left(cth\left(\frac{\beta}{2}\sqrt{\tilde{x}_1}\right)-\frac{1}{sh\left(\frac{\beta}{2}\sqrt{\tilde{x}_1}\right)}\left(ch\left[\left(\frac{\beta}{2}-2\tau_0\right)\sqrt{\tilde{x}_1}\right]-\right.\right.\right.
$$

$$
\left.\left.\left.-ch\left[\left(\frac{\beta}{2}\right)\sqrt{\tilde{x}_1}\right]+ch\left[\left(\frac{\beta}{2}-2\tau_0\right)\sqrt{\tilde{x}_1}\right]\right)\right)-
$$

$$
-\frac{(\omega^2-\tilde{x}_1)}{\sqrt{\tilde{x}_2}}\left(cth\left(\frac{\beta}{2}\sqrt{\tilde{x}_2}\right)-\frac{1}{sh\left(\frac{\beta}{2}\sqrt{\tilde{x}_2}\right)}\left(ch\left[\left(\frac{\beta}{2}-2\tau_0\right)\sqrt{\tilde{x}_2}\right]-ch\left[\left(\frac{\beta}{2}\right)\sqrt{\tilde{x}_2}\right]+\right.\right.
$$

$$
\left.\left.\left.+ch\left[\left(\frac{\beta}{2}-2\tau_0\right)\sqrt{\tilde{x}_2}\right]\right)\right)\right\}, \tag{15}
$$

where $$\tilde{x}_{1,2}=\frac{1}{2}\left(\omega^2+\omega_L{}^2+\frac{C^2}{\omega_L{}^2}\right)\mp\frac{1}{2}\sqrt{\left(\omega^2+\omega_L{}^2+\frac{C^2}{\omega_L{}^2}\right)^2-4\omega^2\omega_L{}^2}$$

$$
\tilde{\gamma}=\sqrt{\left(\omega^2+\omega_L{}^2+\frac{C^2}{\omega_L{}^2}\right)^2-4\omega^2\omega_L{}^2}\,.
$$

Here $C$ is the coefficient of interaction with $\omega_L$. The same expression in the atomic units (Bohr units) yields

$$
S=\frac{1}{2}\frac{E_d}{\hbar}a_d^2\varepsilon_0^{*2}l_1^2\left(\frac{l_2}{2l_1}\tau_0^*-\tau_0^{*2}\varepsilon_T^*-\right.
$$

$$-\frac{1}{2\gamma^*}\left\{\frac{\left(\varepsilon_0^{*2}-x_2^*\right)}{\sqrt{x_1^*}}\left(cth\left(\frac{\sqrt{x_1^*}}{2\varepsilon_T^*}\right)-\frac{1}{sh\left(\frac{\sqrt{x_1^*}}{2\varepsilon_T^*}\right)}\left(2ch\left[\left(\frac{1}{\varepsilon_T^*}-2\tau_0^*\right)\frac{\sqrt{x_1^*}}{2}\right]-ch\left[\frac{\sqrt{x_1^*}}{2\varepsilon_T^*}\right]\right)\right)-\right.$$

$$\left.-\frac{\left(\varepsilon_0^{*2}-x_1^*\right)}{\sqrt{x_2^*}}\left(cth\left(\frac{\sqrt{\tilde{x}_2}}{2\varepsilon_T^*}\right)-\frac{1}{sh\left(\frac{\sqrt{\tilde{x}_2}}{2\varepsilon_T^*}\right)}\left(2ch\left[\left(\frac{1}{\varepsilon_T^*}-2\tau_0^*\right)\frac{\sqrt{x_2^*}}{2}\right]-ch\left[\frac{\sqrt{\tilde{x}_2}}{2\varepsilon_T^*}\right]\right)\right)\right\}, \quad (16)$$

where $x_{1,2}^*=\frac{1}{2}\left(\varepsilon_0^{*2}+\varepsilon_L^{*2}+\frac{\gamma_0^*}{\varepsilon_L^{*2}}\right)\mp\frac{1}{2}\sqrt{\left(\varepsilon_0^{*2}+\varepsilon_L^{*2}+\frac{\gamma_0^*}{\varepsilon_L^{*2}}\right)^2-4\varepsilon_0^{*2}\varepsilon_L^{*2}}$

$$\tilde{\gamma}=\sqrt{\left(\varepsilon_0^{*2}+\varepsilon_L^{*2}+\frac{\gamma_0^*}{\varepsilon_L^{*2}}\right)^2-4\varepsilon_0^{*2}\varepsilon_L^{*2}}\,.$$

$$\tau_0^*=\frac{1}{\varepsilon_0^*}Arcsh\left[\frac{a^*-b^*}{a^*+b^*}sh\frac{\varepsilon_0^*}{2\varepsilon_T^*}\right]+\frac{1}{2\varepsilon_T^*}$$

$$\varepsilon_T^*=\frac{\hbar}{\beta E_d},\ \varepsilon_L^*=\frac{\hbar\omega_L}{E_d},\ \beta=\frac{\hbar}{\varepsilon_T^* E_d},\ \varepsilon_T^{*2}=\frac{4U_0^*}{q_0^{*2}},U_0^*=\frac{U_0}{E_d},\ E_d=\frac{m^*e^4}{2\hbar^2\varepsilon^2}$$

Here $E_d$ is the effective Bohr energy, $m^*$ is the effective mass, $e$ is the electron charge, $\varepsilon$ is the dielectric permeability, $U_0$ is the barrier height, or potential amplitude (see, e.g., Fig. 6),

$$l_1=a^*+b^*,\ l_2=3a^*-b^*,\ a^*=\frac{q_0}{a_d},\ b^*=\frac{q_1}{a_d},\ \gamma_0^*=\frac{\hbar^4C^2}{E_d^4},\ \frac{q_1}{q_0}=\frac{b^*}{a^*}=\frac{b}{a}, \quad (17)$$

$a_d=\frac{\varepsilon\hbar^2}{m^*e^2}$ is the effective Bohr radius, $q_0$ is the quantum dot radius, and $q_1$ is the radius of other quantum dot or semi-width of quantum well.

The results of the numerical calculation for $\Gamma$ in InSb quantum dots are shown in Fig. 2.

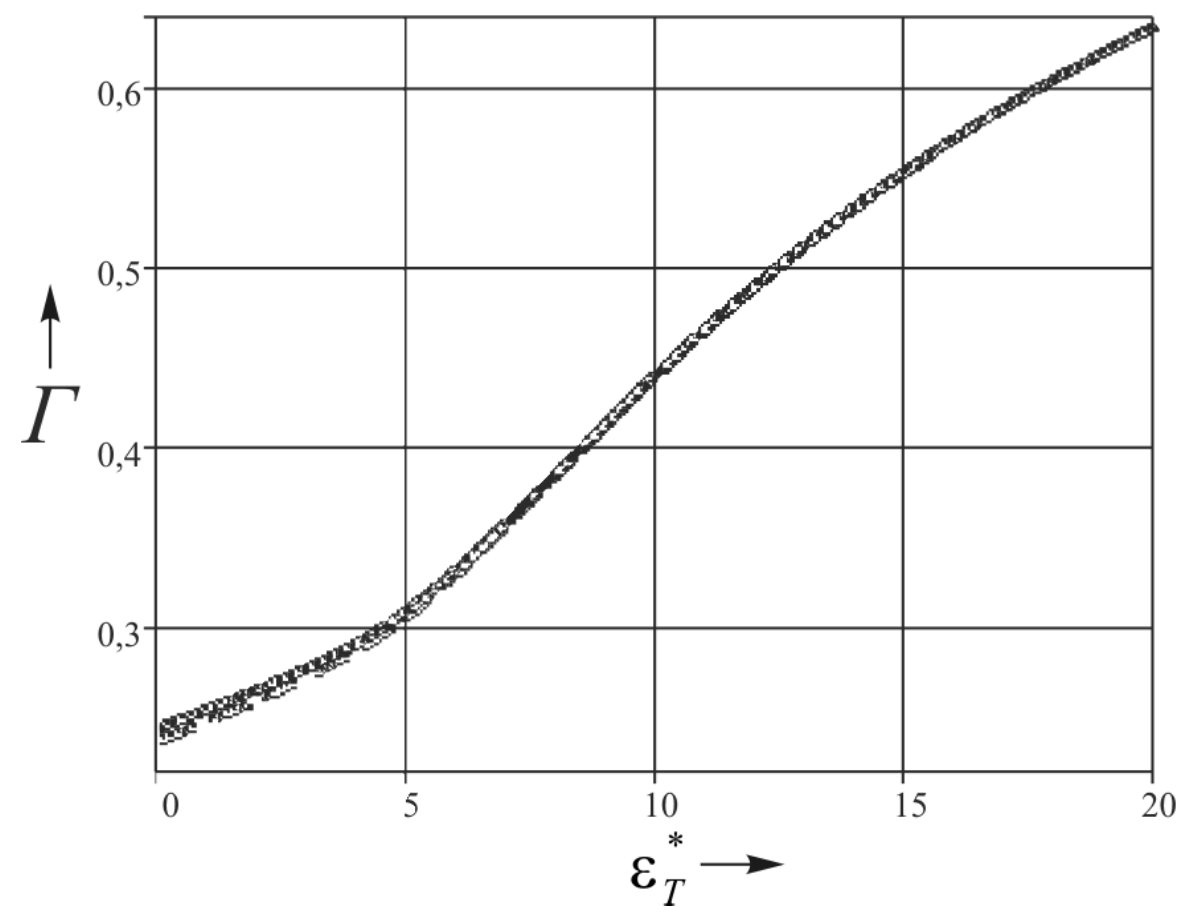

**Fig. 2.** $\varepsilon_T^*$ dependence of $\Gamma$ on in the InSb based quantum dot – quantum well system.

As shown in Figs. 3 and 4, the tunneling probability is very sensitive to both the frequency of local vibration and the coupling constant $C_i$. Indeed, the effective electron vibrational mode coupling increases with the vibration frequency resulting in an increase of the tunneling electron energy and, therefore, enhances the tunneling probability (the transition from curve 1 to curve 2 in Figs. 3 and 4). Fig. 3 reveals some interesting features for quantum molecule systems. If the radius of quantum dot and the width of quantum well are the same, the blockade in the electron tunneling occurs (see the minimum in Fig. 3). A similar dependence was found in the dynamic control of electronic states in double quantum dots with weak dissipation [6]. Furthermore, there are two temperature controlled maxima in the tunneling probability: (a) a quantum dot radius is greater that the semi-width of quantum well (or other quantum dot radius), the first maximum, and (b) a quantum dot radius is lesser than the semi-width of quantum well, the second maximum. The temperature dependence of the right maximum (b) and the left maximum (a) is shown in Fig. 5. When the quantum dot radius is greater than the semi-width of quantum well (or radius of the other quantum dot in quantum molecule), a threshold type behavior takes place as shown in Fig. 4.

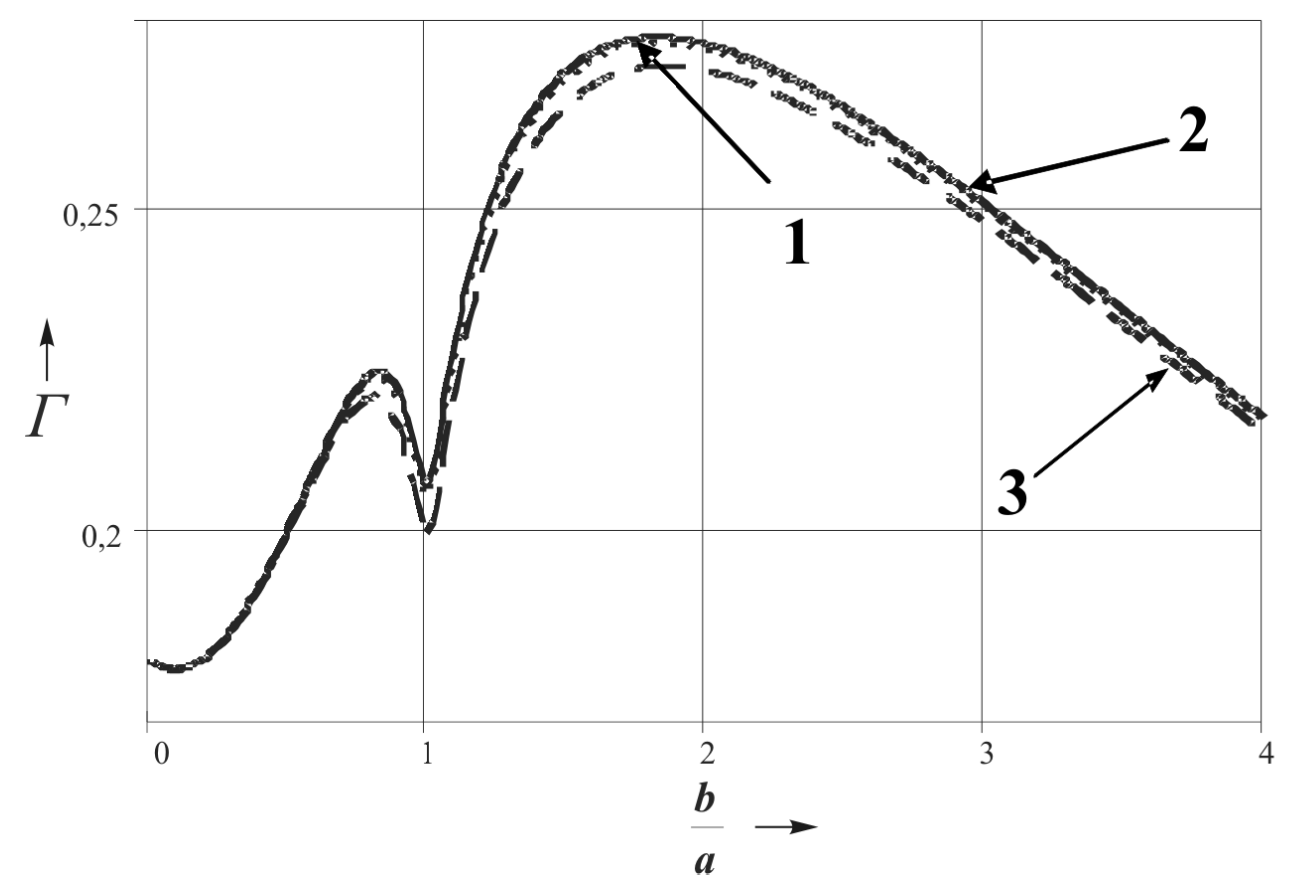

**Fig. 3.** Dependence of $\Gamma$ on the asymmetry ratio parameter *b/a* for quantum molecule (based on InSb): **1.** $U_0^* = 200, \varepsilon_T^* = 3$, $\varepsilon_L^* = 1$, $\gamma_0^* = 10$; **2.** $U_0^* = 200, \varepsilon_T^* = 3$, $\varepsilon_L^* = 10$, $\gamma_0^* = 10$; **3.** $U_0^* = 200, \varepsilon_T^* = 3$, $\varepsilon_L^* = 1$, $\gamma_0^* = 50$.

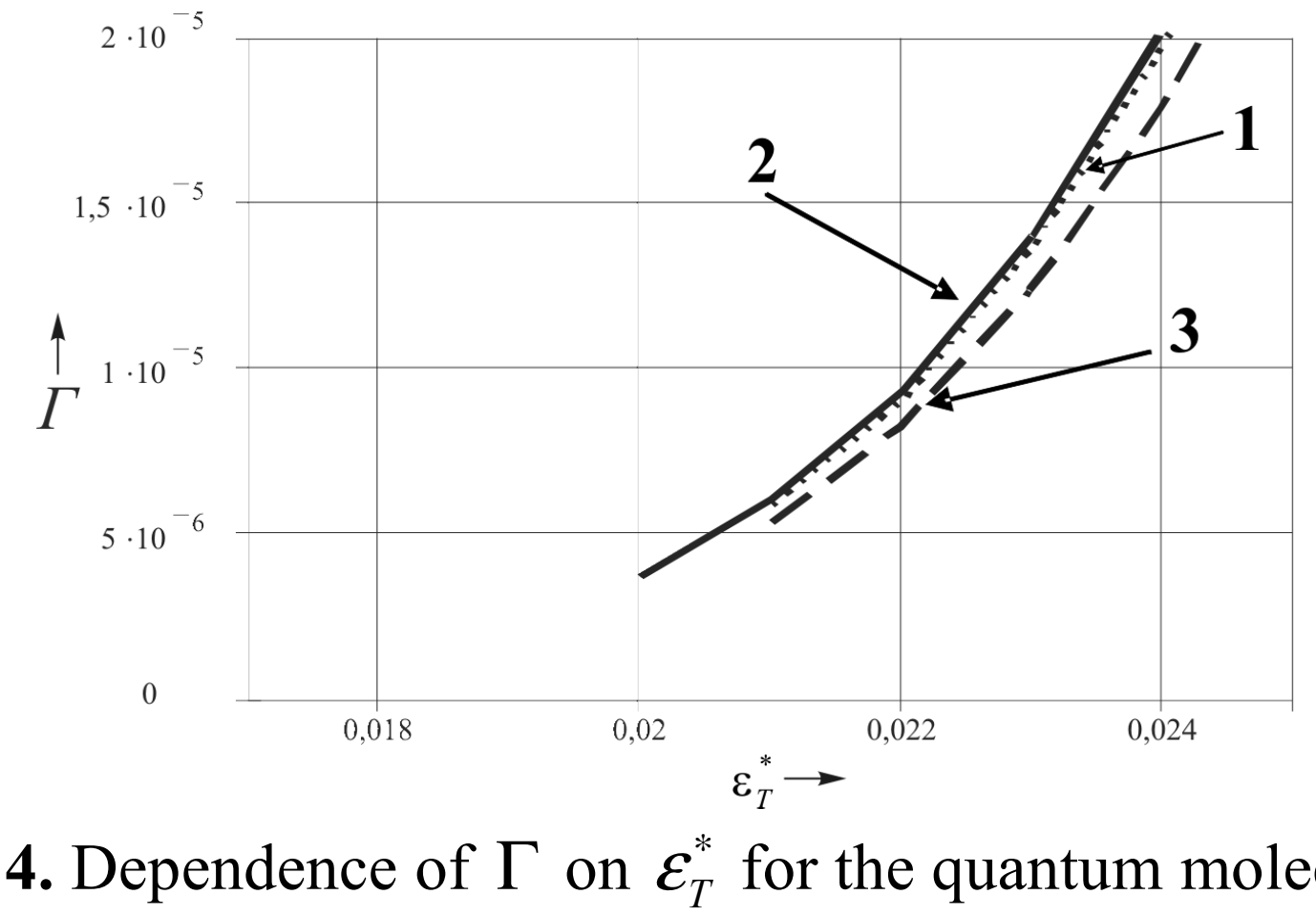

**Fig. 4.** Dependence of $\Gamma$ on $\varepsilon_T^*$ for the quantum molecule (based on InSb), under the condition *b/a*<1: **1.** $U_0^* = 200$, $b/a = 0{,}98$, $\varepsilon_L^* = 1$, $\gamma_0^* = 10$; **2.** $U_0^* = 200$, $b/a = 0{,}98$ , $\varepsilon_L^* = 10$, $\gamma_0^* = 10$; **3.** $U_0^* = 200$, $b/a = 0{,}98$ , $\varepsilon_L^* = 1$, $\gamma_0^* = 50$.

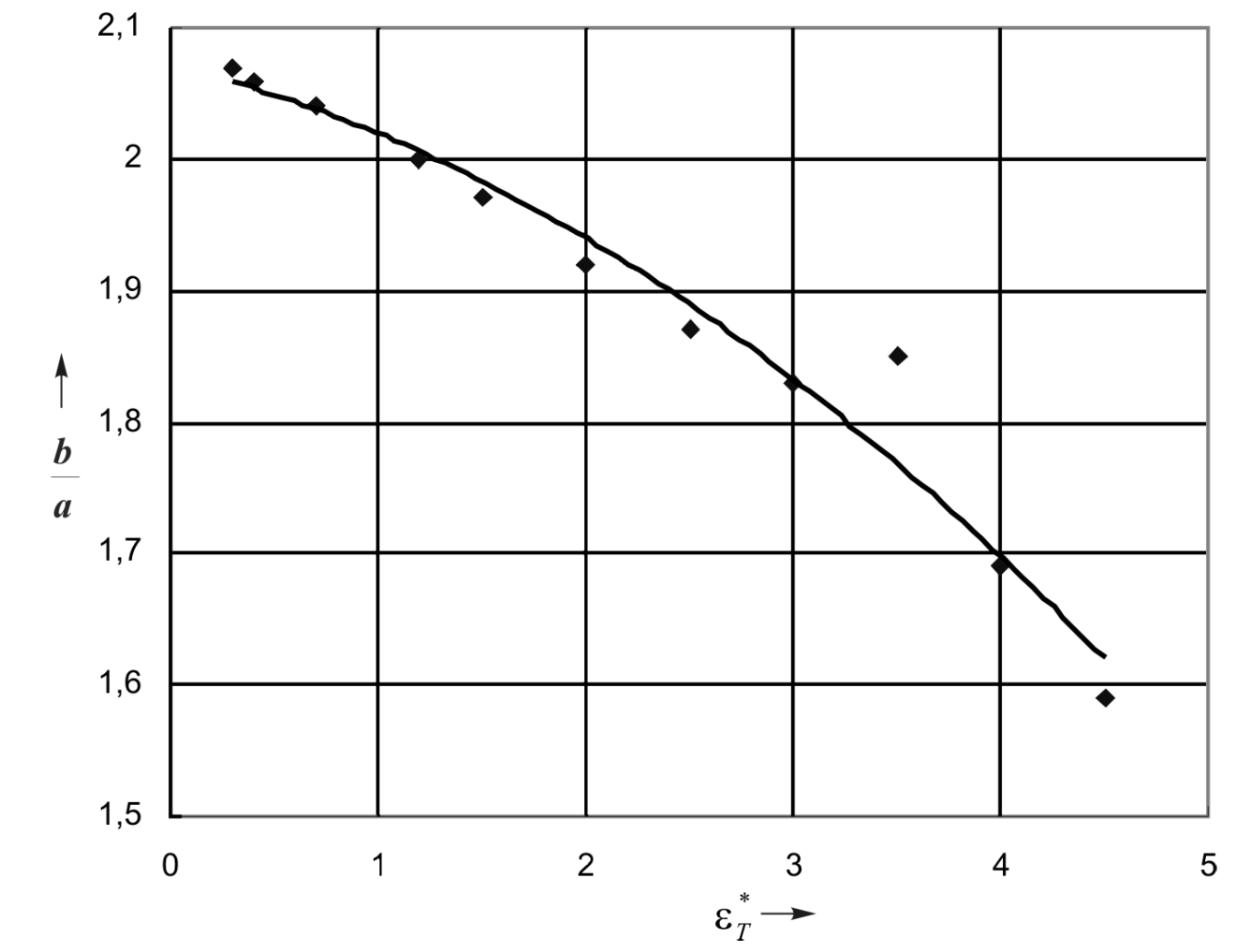

a)

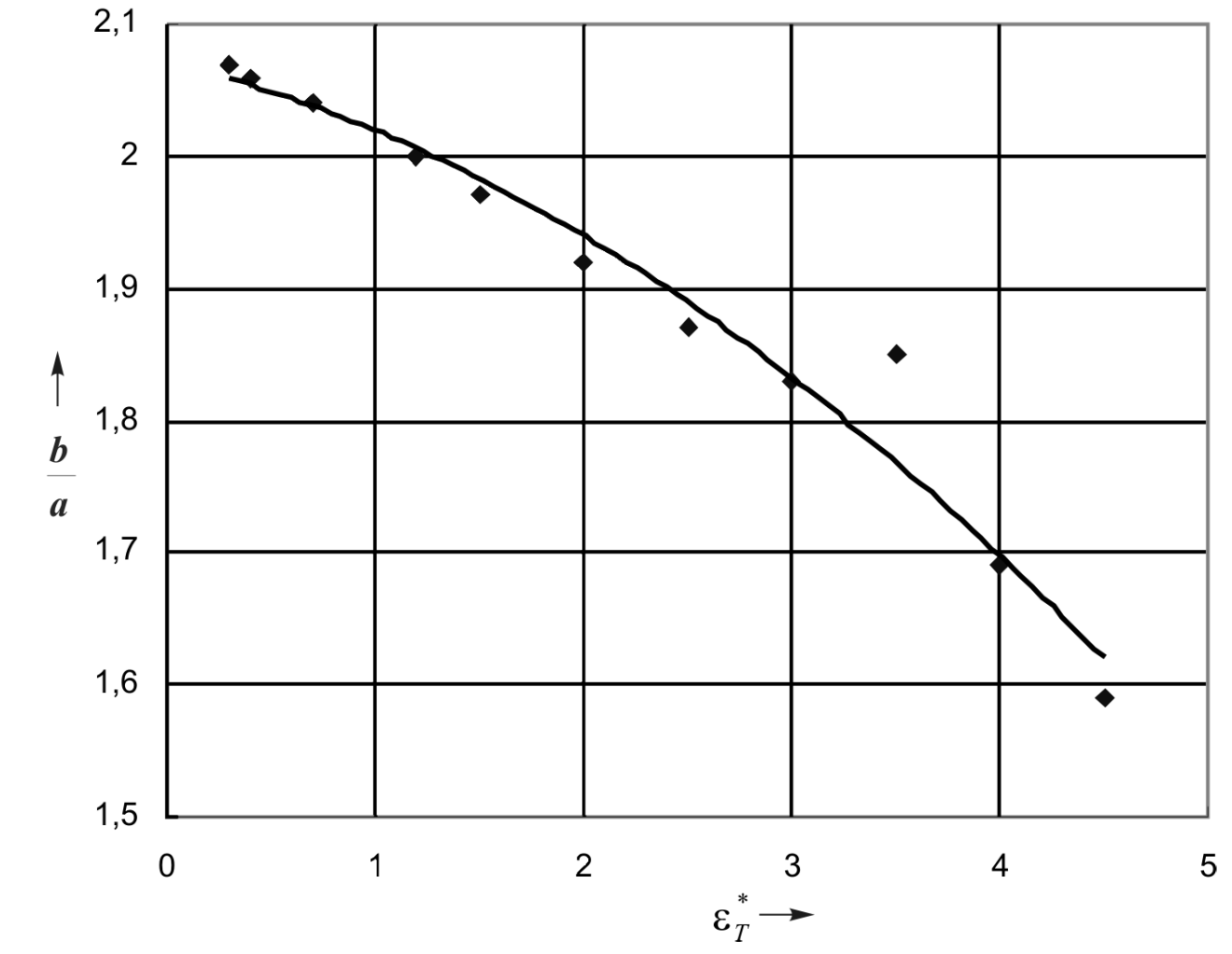

b)

**Fig. 5.** Dependence for extreme ratio values of $b/a$ on $\varepsilon_T^*$ for (a) $b/a > 1$ and (b) $b/a < 1$.

The second tunneling system, quantum dot - bulk contact, is considered in the following Section.

## 3. Quantum dot – bulk contact systems

In a quantum dot - bulk contact tunneling system, a characteristic length of the bulk contact much larger than a quantum dot radius, thus, the limit of $\tau_0 \to 0$ should be taken. Then the Euclidean action becomes

$$S_B = \frac{q_0^2 \beta}{2}\left[\sum_{n=-\infty}^{\infty} \frac{1}{\nu_n^2 + \omega_0^2 + \zeta_n}\right]^{-1} \tag{18}$$

For a single vibrational mode ($\omega_L$), when $\frac{C}{\omega_0^2} << 1$ and $\frac{C}{\omega_L^2} << 1$; $\zeta_n = \frac{C^2 \nu_n^2}{\omega_L^2 (\omega_L^2 + \nu_n^2)}$, Eq. (18) yields

$$S_B = q_0^2 \left[\frac{A}{\sqrt{\tilde{\lambda}_1}} cth\left(\frac{\sqrt{\tilde{\lambda}_1}\,\beta}{2}\right) + \frac{B}{\sqrt{\tilde{\lambda}_2}} cth\left(\frac{\sqrt{\tilde{\lambda}_2}\,\beta}{2}\right)\right]^{-1}, \tag{19}$$

where $A = \frac{\omega_L^2 - \tilde{\lambda}_1}{\tilde{\lambda}_2 - \tilde{\lambda}_1}$, $B = \frac{\omega_L^2 - \tilde{\lambda}_2}{\tilde{\lambda}_1 - \tilde{\lambda}_2}$; и $\tilde{\lambda}_{1,2} = \frac{1}{2}\left[\left(\omega_0^2 + \omega_L^2\right) + \frac{C^2}{\omega_L^2}\right] \pm$
$\pm \frac{1}{2}\sqrt{\left(\omega_0^2 - \omega_L^2\right)^2 + \frac{2C^2}{\omega_L^2}\left(\omega_0^2 + \omega_L^2\right) + \frac{C^4}{\omega_L^4}}$.

Then $\Gamma$ (with an exponential accuracy) is given by

$$\Gamma = \exp\left\{-\frac{M q_0^2}{\hbar}\left[\frac{A}{\sqrt{\tilde{\lambda}_1}} cth\left(\frac{\sqrt{\tilde{\lambda}_1}\,\beta}{2}\right) + \frac{B}{\sqrt{\tilde{\lambda}_2}} cth\left(\frac{\sqrt{\tilde{\lambda}_2}\,\beta}{2}\right)\right]^{-1}\right\} \tag{20}$$

For further convenience, the tunnel probability $\Gamma$ can be rewritten in the atomic units as follows:

$$\Gamma \sim \exp\left\{-\frac{q_0^{*2}}{2}\left[\frac{A^*}{\sqrt{\lambda_1^*}}\,cth\left(\frac{\sqrt{\lambda_1^*}}{2\,\varepsilon_T^*}\right)+\frac{B^*}{\sqrt{\lambda_2^*}}\,cth\left(\frac{\sqrt{\lambda_2^*}}{2\,\varepsilon_T^*}\right)\right]^{-1}\right\}, \tag{21}$$

where $U_0^* = U_0 / E_d$; $q_0^* = q_0 / a_d$; $\varepsilon_L^* = \hbar\omega_L / E_d$; $\gamma_0^* = \hbar^4 c^2 / E_d^4$; $\varepsilon_T^* = k_B T / E_d$; $A^* = \left(\varepsilon_L^{*2} - \lambda_1^*\right)/\left(\lambda_2^* - \lambda_1^*\right)$; $B^* = \left(\varepsilon_L^{*2} - \lambda_2^*\right)/\left(\lambda_1^* - \lambda_2^*\right)$; $\varepsilon_0^{*2} = 4U_0^* / q_0^{*2}$;
where $\lambda_{1,2}^*$ are determined as

$$\lambda_{1,2}^* = \frac{1}{2}\left[\left(\varepsilon_0^{*2}+\varepsilon_L^{*2}\right)+\frac{\gamma_0^*}{\varepsilon_L^{*2}}\right] \pm \frac{1}{2}\sqrt{\left(\varepsilon_0^{*2}-\varepsilon_L^{*2}\right)^2+\frac{2\gamma_0^*}{\varepsilon_L^{*2}}\left(\varepsilon_0^{*2}+\varepsilon_L^{*2}\right)+\frac{\gamma_0^{*2}}{\varepsilon_L^{*4}}}\,. \tag{22}$$

A semiclassical approximation is valid only for quite “wide” potentials when

$$\chi \cdot l >> 1,\ \chi = \sqrt{U_0^* - \varepsilon_0^*}\ / a_d,\ l = q_0 - l_0.$$

From these equations one can estimate the value of $q_0^* \equiv R_0 / a_d$, that satisfies the following inequality:

$$q_0^*\sqrt{U_0^* - \frac{2\sqrt{U_0^*}}{q_0^*}}\left(1-\sqrt{\frac{2\sqrt{U_0^*}}{q_0^*\,U_0^*}}\right) >> 1\,. \tag{23}$$

Consequently, we obtain, that $q_0^* >> 2/\sqrt{U_0^*}$; hence, for the quantum dot, based on InSb, ($a_d \sim 65\,nm$, $E_d \sim 10^{-3}\ eV$, $U_0 = 0.2\,eV$) we have $q_0^* \sim 1.4$ (or in conventional units $q_0 \equiv R_0 \approx 91\,nm$) as shown in Fig. 6.

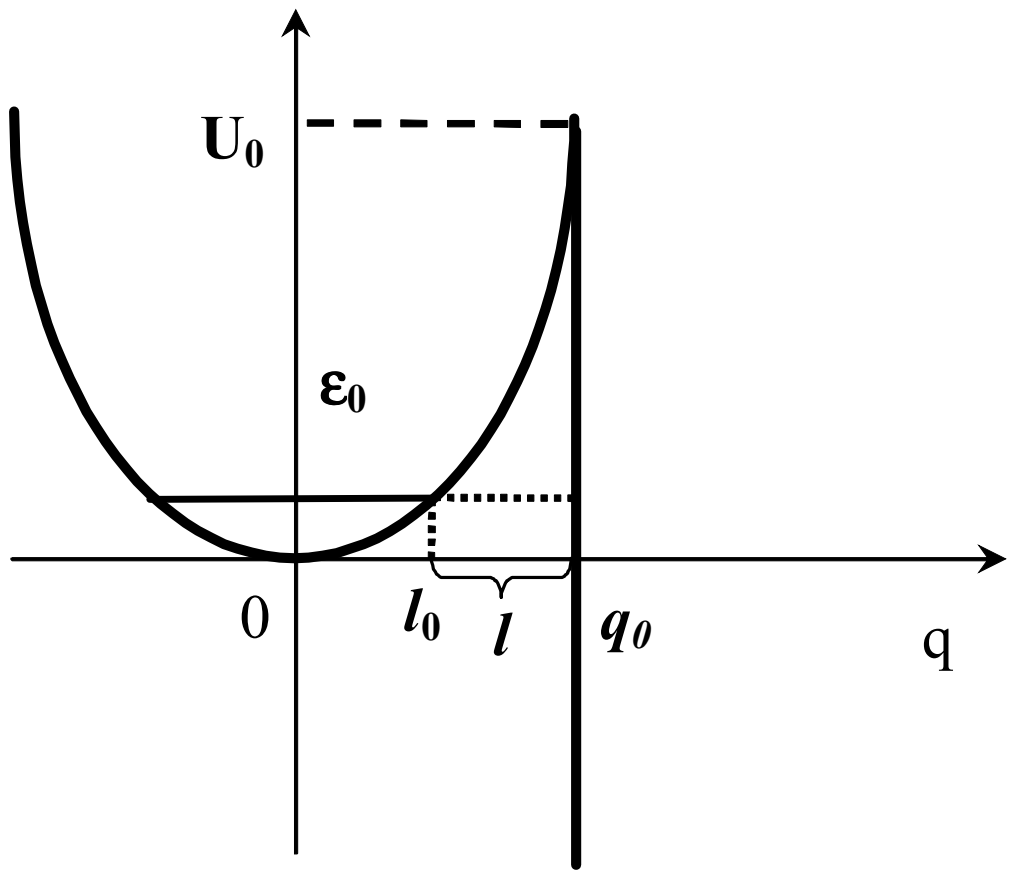

**Fig. 6.** Potential energy for the electron for a "quantum dot – bulk contact" system. Here $\varepsilon_0$ - the QD ground state energy; $l$ - the barrier width; $q_0 \equiv R_0$ is a quantum dot radius; $l_0 = q_0\sqrt{\varepsilon_0 / U_0}$ , and $U_0$ is a quantum dot confinement amplitude or the barrier height).

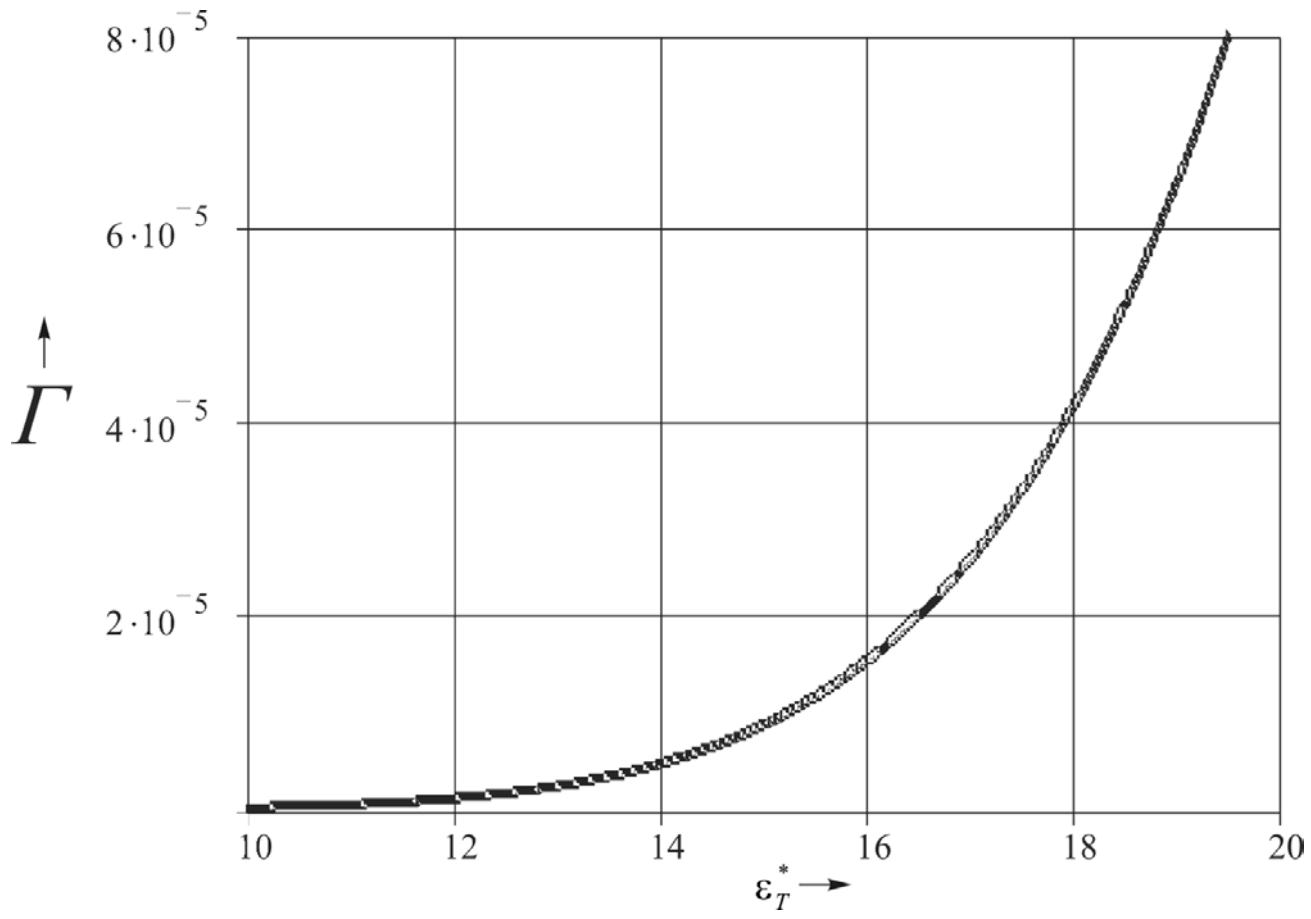

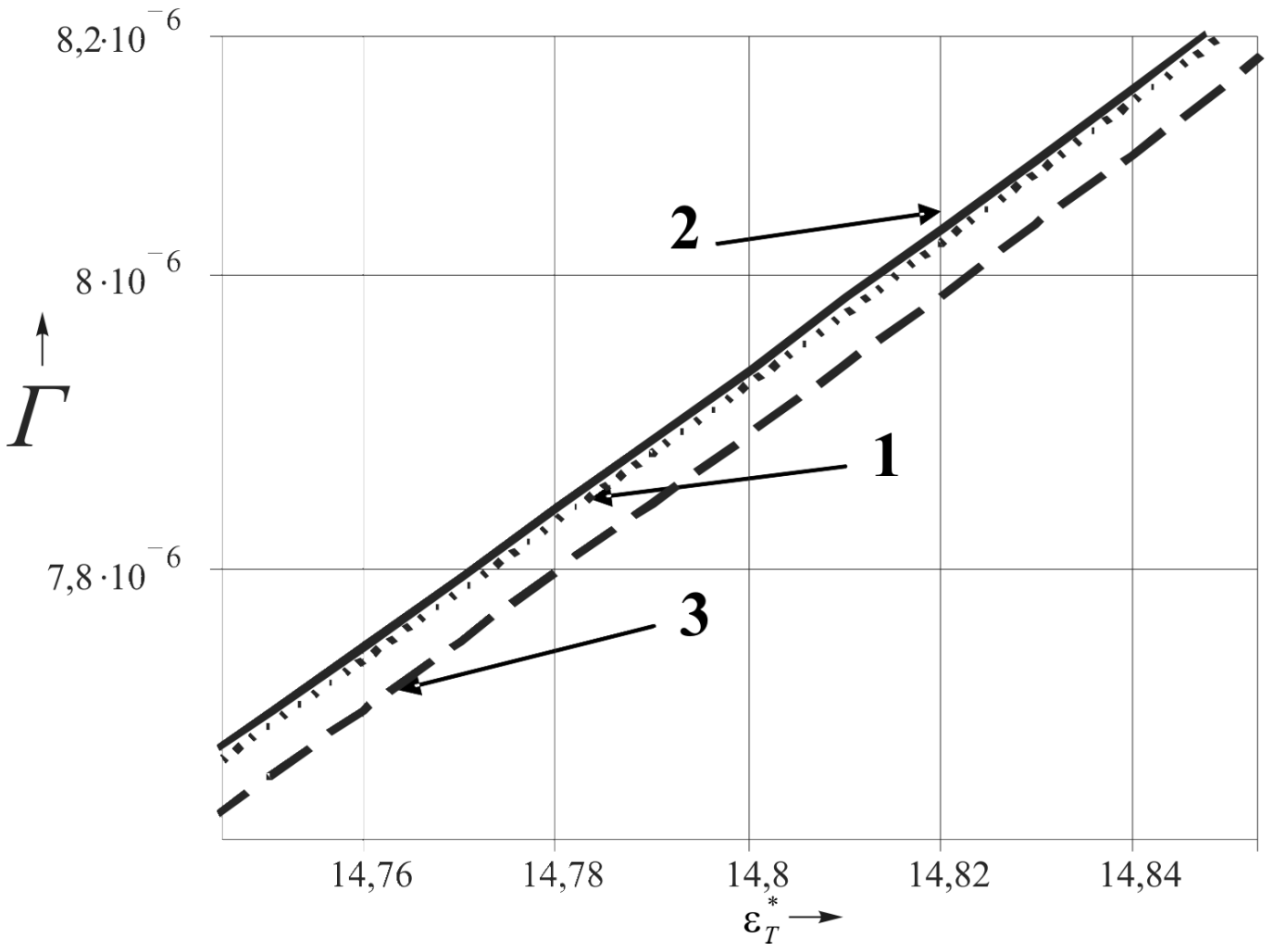

**Fig. 7.** Dependence of $\Gamma$ on $\varepsilon_T^*$ for the quantum dot, based on InSb:

1. $q_0^* = 1.4$, $\varepsilon_L^* = 1$, $\gamma_0^* = 10$, 2 . $q_0^* = 1.4$, $\varepsilon_L^* = 10$, $\gamma_0^* = 10$ and 3. $q_0^* = 1.4$, $\varepsilon_L^* = 1$, $\gamma_0^* = 50$.

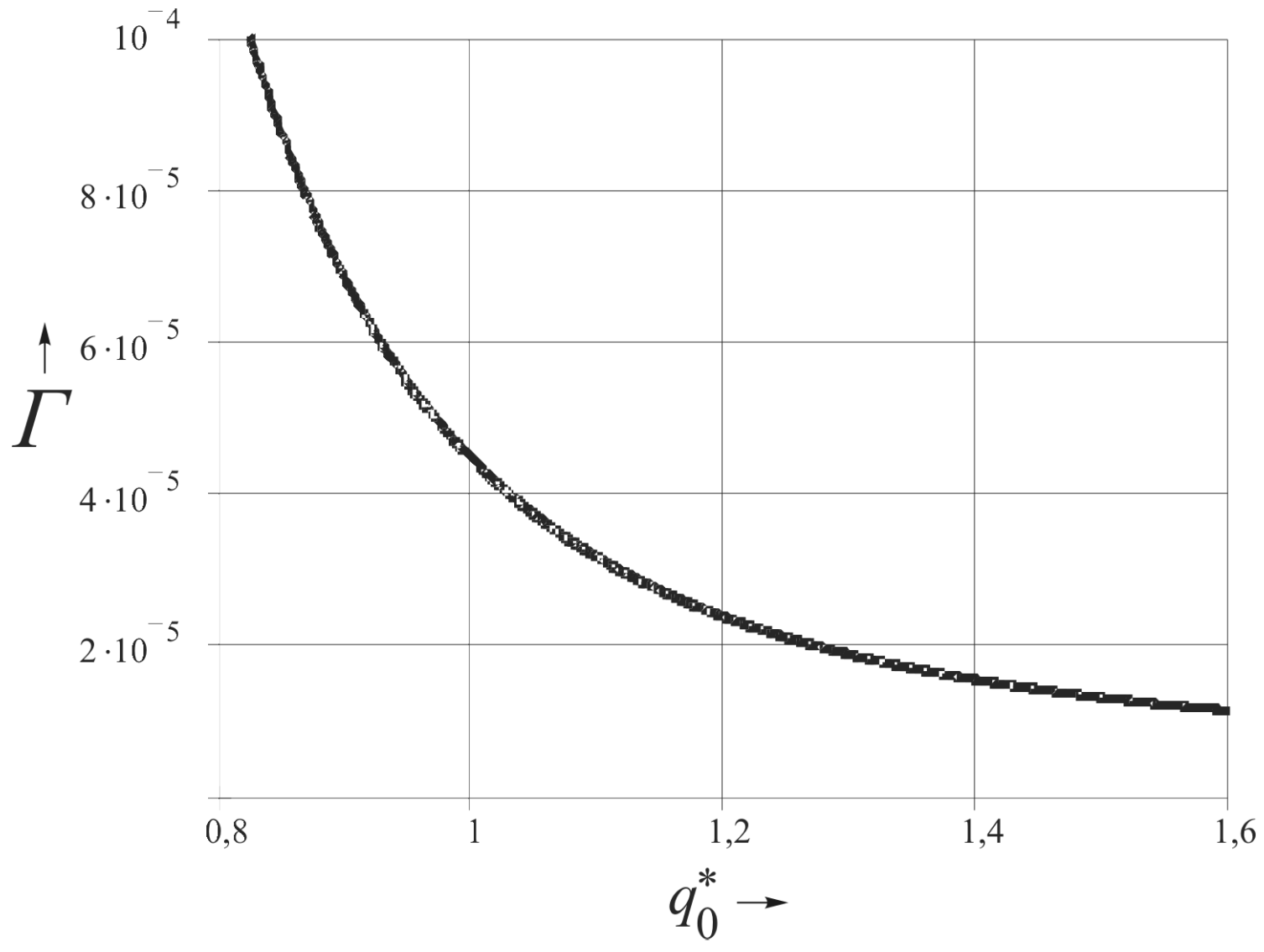

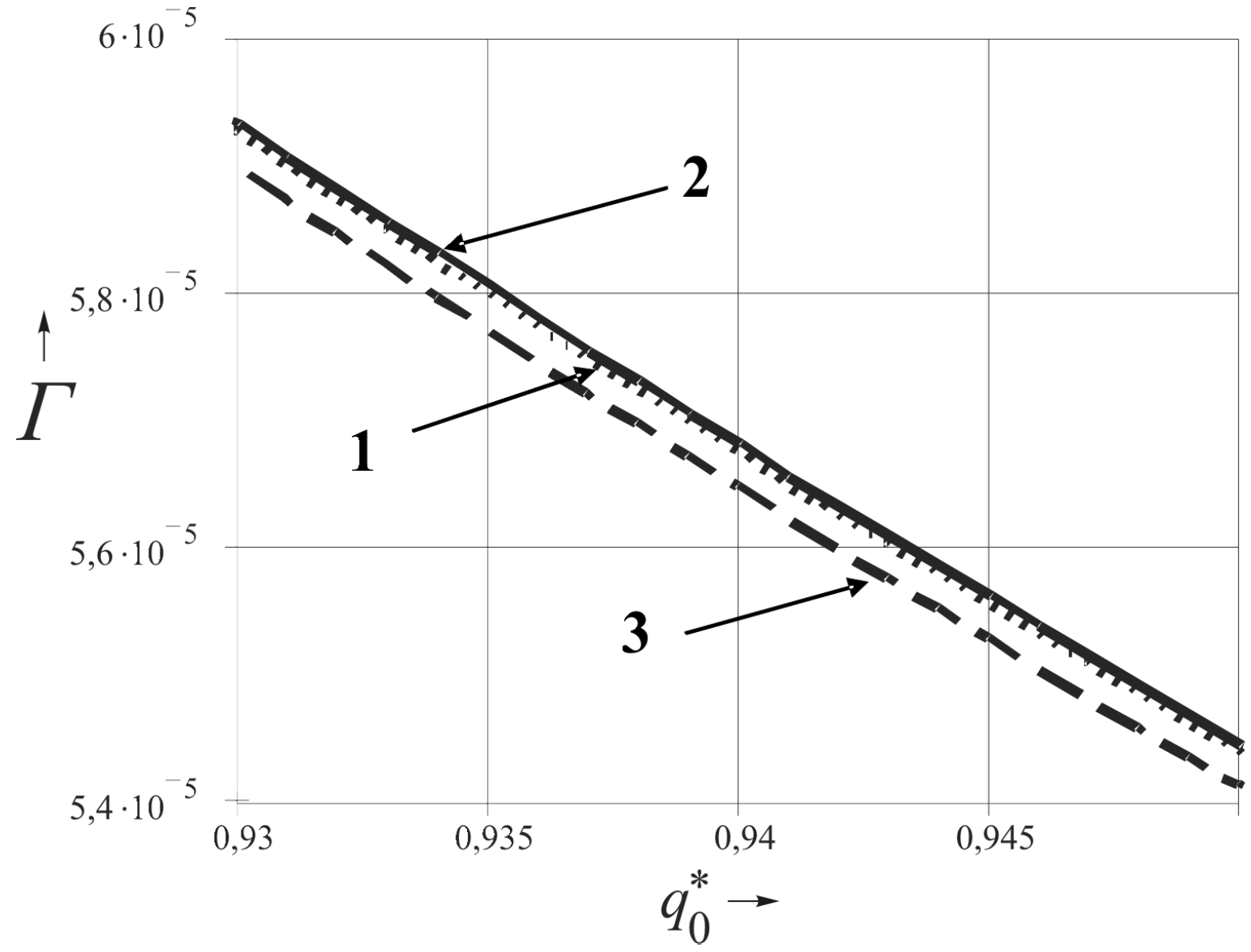

**Fig. 8.** Dependence of $\Gamma$ on $q_0^*$ for the quantum dot, based on InSb:1. $U_0^* = 200, \varepsilon_T^* = 16$, $\varepsilon_L^* = 1$, $\gamma_0^* = 10$, 2. $U_0^* = 200$, $\varepsilon_T^* = 16$, $\varepsilon_L^* = 10$, $\gamma_0^* = 10$ and 3. $U_0^* = 200, \varepsilon_T^* = 16$, $\varepsilon_L^* = 1$, $\gamma_0^* = 50$.

The temperature control for tunneling in “quantum dot - bulk contact” structures has been studied. The tunneling probability dependence on temperature is shown in Fig. 7. As in a double-well potential model (for a quantum molecule), the tunneling probability is sensitive to the local phonon mode frequency and to the coupling constant to the environment (see Figs. 7 and 8). In the same manner, the monotonic decrease of the tunneling probability $\Gamma$ with a quantum dot radius is shown in Fig. 8.

## 4. Summary

The possibility of applicability of the dissipative tunnel theory to the temperature control for tunneling in structures with quantum dots has been demonstrated. The effect of a single electron blockade in the limit of quantum dot has been also revealed in the quantum molecule structures. We hope that the effects described in this paper can be identified in experiments with the use of quantum dots on tip of tunnel microscope.

**Acknowledgement.** This work was partially supported by NSF grant (No ITR-0422690).